# Competing Magnetic States in the Candidate Altermagnet GdAlGe

*Oleg E. Parfenov,[1] Dmitry V. Averyanov,[1] Ivan S. Sokolov,[1] Alexey N. Mihalyuk,[2,3] Ivan A. Yakovlev,[4] Oleg A. Kondratev,[1] Alexander N. Taldenkov,[1] Andrey M. Tokmachev,[1] Vyacheslav G. Storchak[1,\*]*

[1] National Research Center "Kurchatov Institute", Kurchatov Sq. 1, 123182 Moscow, Russia

[2] Institute of High Technologies and Advanced Materials, Far Eastern Federal University, 690950 Vladivostok, Russia

[3] Institute of Automation and Control Processes FEB RAS, 690041 Vladivostok, Russia

[4] Kirensky Institute of Physics, Federal Research Center KSC SB RAS, 660036 Krasnoyarsk, Russia

[\*] Corresponding author; e-mail: vgstorchak9@gmail.com

Altermagnetism, a newly discovered magnetic order, combines zero net magnetization with non-relativistic spin splitting of electronic bands. Its ability to utilize the advantages of both antiferromagnets and ferromagnets is highly promising for spintronic applications. Currently, the merge of altermagnetism and weak ferromagnetism in a single material excites significant interest as it provides additional control mechanisms over material properties. However, the role of dimensionality in this interplay is yet to be explored. Here, we study magnetism and electron transport in epitaxial films of the candidate altermagnet GdAlGe ranging from bulk-like to a single monolayer. The films exhibit the anomalous Hall effect and negative magnetoresistance. In contrast to altermagnetic GdAlSi, the candidate altermagnet GdAlGe demonstrates an admixture of the ferromagnetic state which contribution increases as the system approaches the 2D limit. The coexistence of the magnetic states induces technologically important intrinsic exchange bias. The present work underpins future studies and applications of nanoscale altermagnets.

## 1. Introduction

Altermagnets (AMs) emerge as a promising class of magnetic materials.[1-3] They are capable of revolutionizing the research on fundamental properties of magnetic systems and their applications in spintronics. The hallmark of AMs is a combination of zero net magnetization, as in antiferromagnets (AFMs), and spontaneous non-relativistic spin splitting (NRSS) responsible for spin-based phenomena, as in ferromagnets (FMs). AMs are advertised as taking the best of both worlds to overcome the bottlenecks of AFM and FM applications. A particularly attractive feature is that the spin responses of AMs do not incur any significant stray fields. The functional compatibility of AMs with topological insulators and superconductors is handy for spintronic devices.[4] In studies of AMs, most attention has been drawn to observation of the anomalous Hall effect (AHE)[5-7] but research on other spin-based effects, such as charge-spin interconversion,[8-10] piezomagnetism,[11] the anomalous Nernst effect[12] or the planar Hall effect,[13] is gaining momentum. All those properties stem from specific symmetries of the mutually compensated spin sublattices, compatible with NRSS. The emergence of NRSS in AMs is verified by angle-resolved photoemission spectroscopy, both spin-resolved[14] and spin-integrated.[15] A key direction in the AM research is the design of new candidate AMs employing the toolbox of materials science. In particular, alteration of the crystal structure or chemical composition[16,17] can be a practical approach. A great challenge is to create AM materials which fabrication is scalable because the modern spintronics requires planar nanoscale materials that can be seamlessly integrated with the semiconductor technological platforms.

The push to nanoscale has led to an avalanche of theoretical studies of 2D AM materials. 2D AMs are predicted to host a variety of topological states,[18-20] exhibit quantum spin Hall effect,[21] spin-momentum locked transport,[19] magnetocaloric[22] and piezo[20,23-25] effects. Various candidate 2D AMs have been suggested *via* screening of materials databases[26] or designed employing strains,[27] electric fields,[28] twisting,[29] *etc*. However, their experimental realization is lagging behind because the candidate 2D AMs have rather complex chemical compositions and structures. Moreover, the exact AM symmetry in monolayers (MLs) is likely to be broken by a substrate. A more viable approach would be to scale a functional 3D AM material down to the 2D limit: the exact AM symmetry would be lost but the key advantages of AMs may still be preserved.[30] The necessity to deal with such imperfect AMs raises the question of interplay between magnetic states. In general, thickness-dependent competition between magnetic states is a characteristic feature of many 2D magnets such as $CrI_3$,[31,32] rare-earth metalloxenes,[33-36] $Fe_3GeTe_2$,[37] $MnBi_2Te_4$,[38] trigonal

GdAlSi,[39] or CrSBr.[40] The competition is amenable to external pressure,[31] twisting the layers[32] and chemical modification.[37] Quite recently, the admixture of weak FM in AM materials has attracted a lot of attention. The issue is not limited to 2D materials[19] and interfaces[41] but is important for 3D AMs as well.[42-48] The FM moments can be tiny, as in MnTe,[42] or reaching tenths of Bohr magneton per magnetic atom, as in RuF$_4$.[19] The weak FM provides a way to control the properties of AMs;[42-44] it can also be used to demarcate altermagnets from conventional antiferromagnets.[19,47] What is unclear is the role of dimensionality in the interplay between AM and weak FM, calling for experimental studies of thickness-dependent evolution.

It is not trivial to find an AM system for a study of the thickness-dependent interplay between AM and FM states. Take for example tetragonal GdAlSi, a candidate 3D AM material.[49] Dependence of its properties on the film thickness, down to a single ML (a film with a thickness of 1 unit cell), has been probed recently, with an accent on the chiral anomaly and the emerging spontaneous AHE.[30] However, the uncompensated magnetic moment of 3D GdAlSi is negligible and remains such in the 2D limit despite the AM symmetry breaking – MLs of GdAlSi behave as a 2D compensated ferrimagnet.[50] This is in contrast to other Gd-based materials, including trigonal GdAlSi, which ultrathin films demonstrate rather complex magnetic states.[33-35,39] Although tetragonal GdAlSi does not exhibit a measurable FM admixture, an AM suitable for the study can be designed *via* anionic substitution.[51] Ge-based compounds are often similar to their Si-based counterparts as the two elements are neighbors in group 14 of the periodic table. In particular, GdAlGe has the same tetragonal atomic structure as GdAlSi[52] (see Figure 1a). MAlGe compounds are known to host a number of competing magnetic phases,[53,54] suggesting that GdAlGe may differ from GdAlSi regarding the presence of an FM admixture. As far as we know, GdAlGe is yet to be proposed as an AM candidate. To establish its status, we carried out density functional theory (DFT) calculations (see Methods) of the ground magnetic state of tetragonal GdAlGe (shown in Figure 1a). The magnetic structure of GdAlGe is the same as that of GdAlSi; accordingly, NRSS is present in GdAlGe as well (see Figure 1b). Not only GdAlGe is an AM candidate with expected propensity to magnetic state competition – the natural similarity to GdAlSi suggests that its epitaxial films of any given thickness can be synthesized. It explains our choice of GdAlGe for the study.

Here, we synthesize and study epitaxial films of GdAlGe on the germanium substrate, ranging from bulk-like (90 ML) to a single ML. We demonstrate that the AM state of GdAlGe has an admixture of an FM state, appreciable even in the bulk but increasing as the system approaches the ML limit. The competing magnetic states lead to appearance of an intrinsic exchange bias, an important technological property. In electron transport measurements, we explore the evolution of characteristic properties of magnetic materials – the spontaneous AHE and negative magnetoresistance (MR). The most significant changes are detected in the range below 4 ML where a metal-insulator transition (MIT) takes place, accompanied, for example, by a dramatic increase of the negative MR.

## 2. Results and Discussion

### 2.1. Synthesis and Structural Characterization

Synthesis of epitaxial GdAlGe films with the tetragonal structure (Figure 1a) is no small feat. First, the system Gd-Al-Ge contains well more than a dozen of phases of various stoichiometries, from binary $GdGe_x$ and $GdAl_x$ to ternary $Gd_2AlGe_2$, $GdAl_2Ge_2$, $Gd_2Al_3Ge_4$, *etc*. Accordingly, special attention should be paid to matching the amounts of the reactants and the synthesis conditions. Moreover, the GdAlGe stoichiometry supports at least two structures (tetragonal and orthorhombic[52]), probably three – the example of GdAlSi shows that the trigonal structure can be also stabilized; it may even be the most stable one in ultrathin films.[55] Therefore, one should be extremely careful in choosing the substrate as that can stabilize one or another GdAlGe polymorph. In the case of GdAlSi, synthesis of the tetragonal AM polymorph is assisted by the (001) surface of silicon.[30] Similarly, we employ the (001) surface of germanium to stabilize the tetragonal polymorph of GdAlGe. The key reason for the stabilization is the symmetry and lattice match (the mismatch is about 5 %) between the *ab*-face of GdAlGe and the Ge surface. Synthesis of GdAlGe on germanium carries significant advantages: the film becomes directly integrated with a major technological semiconductor and the Ge substrate can be employed as one of the reactants to facilitate the reach of the required stoichiometry by reducing the number of independent reactants. Our aim is to study the dependence of GdAlGe properties on the number of MLs. Therefore, a technique that would ensure a layer-precision growth of the material is needed. Our choice is molecular beam epitaxy (MBE) because it has proven useful in synthesis of GdAlSi[30,39,55] as well as Gd-based 2D materials on germanium.[34,35]

Actually, we used two types of substrates. First, pristine Ge(001) wafers were employed to synthesize rather thick GdAlGe films. However, this type of the substrate turned out to be impractical for ultrathin films (4 ML and thinner). The reason is that the conductivity of the Ge wafer is comparable to the conductivity of ultrathin GdAlGe films, *i.e.* the substrate affects the results of electron transport measurements making their interpretation difficult. To avoid the problem, the ultrathin films of GdAlGe were synthesized on another type of substrate, thin films (7-9 nm) of Ge(001) on high-ohmic Si(001). In both cases, the synthesis started with formation of the bare Ge(001) surface by removal of the natural surface oxide (see Methods). Then, matching amounts of Gd and Al were co-deposited on Ge(001). The optimized substrate temperature for the synthesis was 190 C°; at higher temperatures, the epitaxy is broken. The temperature is significantly lower than that for synthesis of GdAlSi films (400 °C[30]). The GdAlGe film thickness was determined by the metal deposition time. We produced a set of GdAlGe films – rather thick films, 45 ML and 90 ML, emulating the bulk, ultrathin films ranging from 1 ML to 4 ML, and a couple of films of intermediate thickness, 10 ML and 17 ML. GdAlGe is prone to oxidation by air. Therefore, all the films were capped with a layer of $SiO_x$, an amorphous nonmagnetic insulator that does not affect the results of our *ex situ* studies of the GdAlGe films.

The atomic structure of the GdAlGe films was studied by diffraction techniques. First, the state of the sample surface was monitored using reflection high-energy electron diffraction (RHEED) in the growth chamber. A typical 3D RHEED image of 1 ML GdAlGe is shown in Figure 2a. Characteristic RHEED images for some other films are provided in Figure S1 of the Supporting Information. All the RHEED patterns correspond to epitaxial films of the tetragonal polymorph of GdAlGe. In particular, the directions [001] of GdAlGe and Ge coincide. The lattice parameter $a$ = 4.20(4) Å is slightly above the value determined for polycrystalline bulk GdAlGe (4.1521(9) Å[52]). The *ex situ* study of the atomic structure was carried out by X-ray diffraction (XRD) techniques. Figure 2b shows a θ-2θ scan of the bulk-like 90 ML film of GdAlGe. Characteristic XRD scans for some other films are provided in Figure S2 of the Supporting Information. The XRD scans exhibit a series of GdAlGe peaks without any traces of side phases or crystallite orientations. XRD confirms the epitaxial relationship between GdAlGe and Ge established by RHEED. The lattice parameter $c$ = 14.38(4) Å determined for 90 ML GdAlGe agrees with the value 14.415(7) Å[52] for the bulk. It should be noted here that the nominal values of the film thickness should be taken with a grain of salt because they are based on the amounts of deposited Gd and Al. They can be considered as some average values – Figure S1 of the Supporting Information suggests that the film surface can be rather rough; moreover, the roughness increases as the film becomes thicker.

**2.2. Magnetism and Electron Transport in Bulk GdAlGe**

Before studying thickness dependence of GdAlGe properties, it is essential to establish those of bulk GdAlGe. Accordingly, we first consider bulk-like films of GdAlGe. For a candidate AM material, magnetic properties are of particular importance. Figure 3a demonstrates temperature dependence of the molar magnetic susceptibility $\chi_m$ in 90 ML GdAlGe. The general form of the dependence points at an AFM (or AM) transition around 20 K. The anisotropy of the magnetic susceptibility is not significant (see Figure S3 of the Supporting Information). The fit of the $\chi_m(T)$ dependence in the paramagnetic region by the Curie-Weiss law (see Figure S4 of the Supporting Information) determines the Weiss constant $\theta$ = -8.5 K. The sign of $\theta$ suggests the predominance of AFM correlations but the absolute value is rather low, which may point at the presence of FM coupling, as in ultrathin $GdX_2$ (X = Si, Ge) and $EuGe_2$ (isoelectronic to GdAlGe) that exhibit competing magnetic states.[35]

Indeed, magnetization measurements reveal properties characteristic to the FM state. In particular, the *M-H* dependence (Figure 3b) demonstrates a non-linear contribution that we attribute to the FM state. Studies of magnetic field dependence of the FM moment detect a hysteretic behavior (inset in Figure 3b). The total FM moments at 2 K are about 0.13 $\mu_B$/Gd, *i.e.* only 2 % of the magnetic moments expected for the fully polarized FM phase. This estimate agrees with the magnetization measurements of polycrystalline bulk GdAlGe.[56] It points at a relatively small admixture of the FM state to the predominant AM state. The GdAlGe film demonstrates a significant remnant moment (Figure 3c). This magnetic behavior is quite different from that of GdAlSi.[30] To analyze the magnetic structure of GdAlGe, we consider DFT calculated relative energies of GdAlGe with characteristic magnetic configurations (see Figure S5 and Table S1 of the Supporting Information). The relative energy of the FM (full spin polarization) state with respect to the ground AM state is rather high (the same is true for GdAlSi). However, the relative energy of the ferrimagnetic state (per formula unit) is less than 1 meV. Such a state may take the role of the FM state detected in the experiments. In the case of GdAlSi, the energy of the ferrimagnetic state is noticeably higher, 2.3 meV; it may explain the difference in the magnetic behavior of GdAlGe and GdAlSi.

The presence of two magnetic states can give rise to exchange bias (EB), a technologically important effect making an operating principle of magnetic memory technologies.[57] AM-based systems with EB are predicted to generate exotic electric-field induced phenomena.[58] EB is commonly detected in heterostructures combining different magnetic orders. Coupling between coexisting magnetic states in a single material is a source of *intrinsic* EB, detected in Gd and Eu metalloxenes,[59,60] $Fe_xNbS_2$,[61] NiO,[62] and $MnSb_2Te_4$.[63] In the rare-earth metalloxenes, the effect is limited to films of intermediate thickness. Similarly, in GdAlGe, EB is best detected in the 17 ML film. Figure 3d shows a shift of the hysteresis loop along the magnetic field axis. The size of the effect is about the same as in EuX$_2$[60] but significantly lower than that in GdSi$_2$.[59] Nevertheless, it provides direct evidence of coexisting magnetic states.

The magnetic states affect electron transport properties of GdAlGe. Figure 4a demonstrates temperature dependence of resistance for 45 ML GdAlGe. The material is a metal – metallic AMs are suggested to carry significant advantages for applications in spintronics.[64] In a magnetic field, the maximum of $R_{xx}(T)$ shifts to a lower temperature. It suggests that the predominant magnetic order in GdAlGe is AFM (or AM), in agreement with the magnetization measurements. The system exhibits a significant anisotropy of resistance for in-plane magnetic fields regarding their orientation to the electric current (Figure S6 of the Supporting Information); the anisotropy emerges at the magnetic transition temperature. MR in GdAlGe is highly anisotropic as well – its sign depends on the magnetic field direction (Figure S7 of the Supporting Information). Moreover, MR exhibits a significant hysteresis. The main signature of the FM state is the presence of an AHE that emerges below the magnetic transition temperature (Figure 4b). In principle, the AM state can also generate the AHE. However, the magnetic point group of the AM state of GdAlGe, $4'm'm$, does not allow for observation of the AHE in these experiments – the situation is similar to that in GdAlSi where AHE is absent in bulk-like films.[30] Therefore, we attribute the AHE to the admixture of an FM magnetic state.

**2.3. 2D Limit of GdAlGe**

Both magnetic and electron transport properties depend on the film thickness. This dependence comes from different sources – one of them is change of the magnetic state. The FM admixture in GdAlGe films is present irrespective of the film thickness as witnessed by remnant moments (see Figure S8 of the Supporting Information). Moreover, as the film becomes thinner, the FM contribution increases: Figure S9 of the Supporting Information shows that the *M-H* dependence in ultrathin films becomes progressively more FM-like. However, the FM moments are still far from the maximal theoretical value of 7 $\mu_B$/Gd. Comparison of temperature dependences of the FM moments in thick and ultrathin films (Figure S10 of the Supporting Information) suggests that the effect of the film thickness on the magnetic transition temperature is rather weak.

Electron transport in GdAlGe films is sensitive to the film thickness. Figure 5a shows that the conductivity of GdAlGe is rather stable down to 3 ML films. At lower thickness, MIT takes place – in 1 ML GdAlGe, the conductivity drops by more than 2 orders of magnitude. This behavior differs from that in GdAlSi films, both tetragonal[30] and trigonal,[65] which remain metallic down to a single ML. Other transport properties change as well. For instance, the negative MR in 1 ML is 20 times higher than that in 45 ML GdAlGe and reaches about -30 % in an out-of-plane field of 9 T (Figure 5b). The Hall conductivity in thick GdAlGe films at 50 K (above the magnetic transition temperature) is electron-like; it remains about the same down to 4 ML (Figure 5c). However, in 2 ML, the value of $\rho_{xy}$ triples, and in 1 ML, it even changes its sign. Basically, the thickness dependence of electron transport boils down to 2 regimes: (i) weak dependence between 3-4 ML and the bulk; (ii) strong dependence at the MIT below 3 ML. To illustrate the stability of the properties in regime (i), we consider hysteresis in the AHE resistance (Figure 5d) – the loops measured in 4 ML and 90 ML GdAlGe almost coincide.

In general, the 4 ML film exhibits transport properties similar to those of the bulk. In particular, magnetic fields shift the magnetic transition towards lower temperatures (Figure S11 of the Supporting Information); it points at the predominant role of the AFM (or AM) state. Temperature dependence of sheet resistance in-plane magnetic fields demonstrates a significant anisotropy emerging at the magnetic transition temperature (Figure S12 of the Supporting Information). MR in 4 ML GdAlGe exhibits a hysteresis in low in-plane magnetic fields (Figure S13 of the Supporting Information). Low-temperature MR is negative in both in-plane and out-of-plane magnetic fields; its absolute value decreases at elevated temperatures (Figure S14 of the Supporting Information). The MR anisotropy depends on temperature: being noticeable at 2 K, it disappears at 50 K, above the magnetic transition temperature (Figure S15 of the Supporting Information). Also, the film exhibits an AHE resistance that dwindles as temperature increases (Figure S16 of the Supporting Information). In ultrathin films of GdAlGe, the AHE can be caused by both AM and FM states.

The electron transport undergoes significant changes as the film is thinned beyond the MIT, *i.e.* in 2 ML and 1 ML GdAlGe. The temperature dependence of resistivity in the ultrathin films is analyzed within the Mott variable-range hopping model $\rho_{xx} \propto \exp(T^{-\beta})$ where $\beta$ depends on the dimensionality $d$ as $\beta = 1/(d + 1)$. In the case of 2 ML GdAlGe, $\beta = 1/4$ (Figure 6a) that corresponds to 3D strong localization. In contrast, $\beta = 1/3$ in 1 ML GdAlGe (Figure 6b), *i.e.* corresponds to 2D strong localization. Similarly, MIT causing 3D and 2D strong localization in 2 ML and 1 ML, respectively, has been reported for SrSi$_2$.[66] It highlights an important role of each ML in the localization effects. Both 1 ML and 2 ML GdAlGe demonstrate negative MR at various temperatures (Figure S17 of the Supporting Information). The difference is in the magnetic anisotropy: in 1 ML, the MR anisotropy is rather weak but it is practically absent in 2 ML GdAlGe (Figure S18 of the Supporting

Information). As for AHE, we were able to measure it only in 2 ML GdAlGe (Figure 6c). The hopping electron transport in magnetic semiconductors is expected to exhibit the power-law scaling relationship $\sigma_{xy} \propto (\sigma_{xx})^{1.6-1.8}$.[67] This relationship, thought to be universal, is established for many 3D magnets but its applicability to 2D systems is questionable. A power-law scaling was reported for the 2D FM material $Fe_5GeTe_2$ but the exponent (1.4) was too low.[68] In contrast, 2 ML GdAlGe exhibits a scaling relation with an exponent 1.61 (Figure 6d) although the spans of $\sigma_{xx}$ and $\sigma_{xy}$ are relatively small.

## 3. Conclusion

AMs, a class of materials touted as capable of revolutionizing spintronics, can overcome serious problems faced by FMs and traditional AFMs. However, the rapidly growing amount of AM research is yet to provide breakthrough applications. One of the reasons is that real-life applications require nanoscale materials integrated with mainstream technological platforms. Thus, experimental studies of the 2D limit of AMs are in demand despite being very challenging. A particular issue we are interested in is the competition between AM and weak FM states that is currently attracting much attention. We meet the challenge by epitaxial synthesis and studies of GdAlGe films on germanium ranging from 1 to 90 ML. This candidate AM is shown to exhibit an admixture of the FM state that evolves with the film thickness. Our complex study of magnetism and electron transport in GdAlGe films reveals signatures of both magnetic states. In particular, the detected intrinsic EB is a hallmark of the coexisting states. GdAlGe is compared to its close AM analogue, GdAlSi. Their behavior is quite different: the former exhibits coexisting magnetic states whereas the latter does not, even in the ML limit. Moreover, unlike GdAlSi, GdAlGe undergoes a MIT that changes its properties profoundly. It shows that small variations in the AM material constitution, such as the Si/Ge substitution, matter. *De facto*, GdAlGe and GdAlSi belong to different groups of AM materials with respect to the AM/FM competition. It applies to both bulk and ML limits of the materials. Hopefully, the present study will facilitate the development of AM-based ultracompact spintronics.

## 4. Methods

*Synthesis*: GdAlGe films were fabricated in a Riber Compact MBE system under UHV conditions (the base pressure in the growth chamber was kept below $10^{-10}$ Torr). Two types of substrates were employed. One was a square wafer of Ge with a lateral size 1 inch whereas the other was a thin layer of Ge (7-9 nm) on a 5 mm × 10 mm wafer of Si. To produce the tetragonal polymorph, the faces of both substrates were (001), with a miscut angle less than 0.5°. The Ge wafers were wet-etched in 5% $NH_3$. To produce the thin layer of Ge on Si, the Si surface was prepared by wet-etching and annealing in vacuum at 600 °C for 2 hours, followed by heating to 900 °C to get the characteristic 1 × 2 reconstruction. Then, Ge was deposited at 380 °C with a rate 0.25 nm/min. Before the GdAlGe synthesis, both types of the substrates were annealed at 600 °C. The substrate temperature was controlled by a thermocouple, calibrated by a PhotriX ML-AAPX/090 infrared pyrometer. Synthesis of GdAlGe by co-deposition of Al and Gd on the substrate was carried out at 190 °C. 5N Al and 4N Gd were supplied from Knudsen cell effusion sources heated to 905 °C and 1210 °C, respectively. According to a Bayard-Alpert ionization gauge, the pressures of Al and Gd were $6·10^{-9}$ Torr and $1·10^{-8}$ Torr, respectively (the same as in synthesis of GdAlSi[30]). Those fluxes provided the growth of 90 ML GdAlGe in 160 min. The films were protected from air by a 200-nm layer of $SiO_x$ deposited at room temperature.

*Characterization*: The surface of the films in the growth chamber was continuously monitored by a RHEED diffractometer furnished with the kSA 400 analytical system. *Ex situ* characterization of the GdAlGe atomic structure was based on θ-2θ XRD scans recorded in a Rigaku SmartLab 9 kW XRD diffractometer operating at the $CuK_{α1}$ wavelength. Magnetic properties of the GdAlGe films were measured by an MPMS XL-7 SQUID magnetometer employing the reciprocating sample option. Square samples with a lateral size of 5 mm were oriented with respect to external magnetic fields with an accuracy of better than 2°. The lateral electron transport properties of the GdAlGe films were determined by a Lake Shore 9709A measurement system. Four-contact measurements were carried out for samples with a lateral size of 5 mm and adhered to the ASTM standard F76. The ohmic electrical contacts were fabricated by deposition of an Ag-Sn-Ga alloy; their quality was attested by I-V characteristic curves.

*Computational Techniques*: The electronic structure calculations of GdAlGe and GdAlSi were carried out based on the DFT + U approach, as implemented in the Vienna ab initio simulation package (VASP).[69] The projector augmented wave approach[70] was used to describe the electron-ion interaction. The band structure and the relative energies of the magnetic configurations were produced employing the meta-GGA approximation (MS1 functional).[71] A Hubbard U correction[72] of 7 eV was applied to Gd. The kinetic energy cutoff was set to 350 eV; the Brillouin zones of GdAlGe and GdAlSi were sampled with 12×12×6 Γ-centered *k*-point meshes; spin-orbit coupling was taken into account. The lattice constants of the compounds were relaxed to produce $a$(GdAlGe) = 4.16 Å, $c$(GdAlGe) = 14.63 Å, $a$(GdAlSi) = 4.13 Å, and $c$(GdAlSi) = 14.40 Å.

**Acknowledgements**

This work was supported by the Russian Science Foundation [grant No. 24-19-00038].

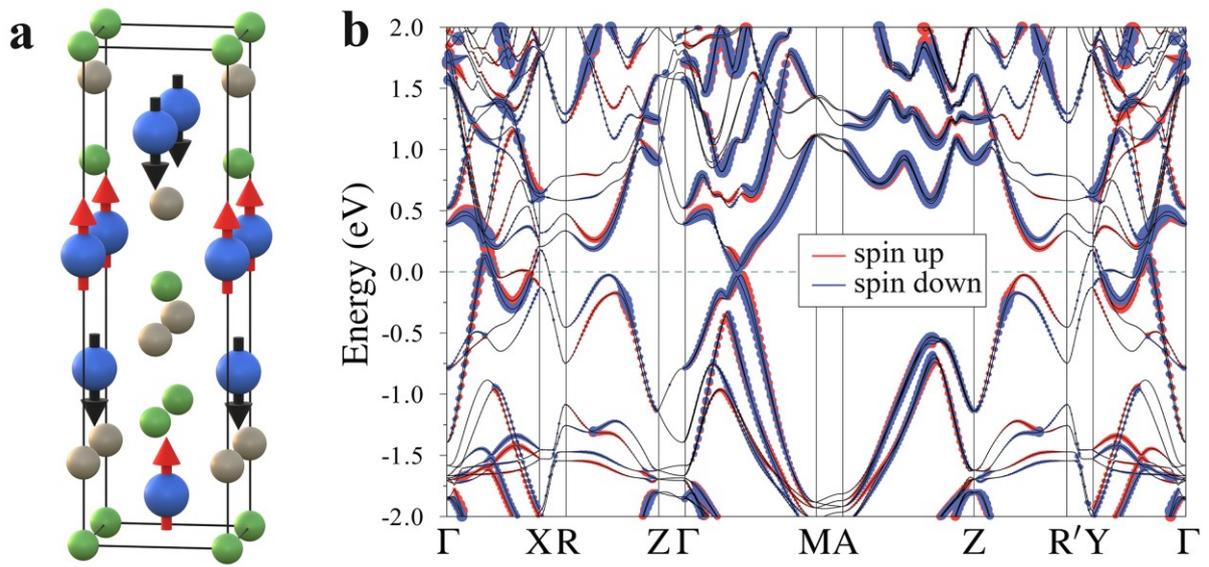

**Figure 1.** Atomic, magnetic, and electronic structures of GdAlGe. a) The unit cell of GdAlGe (Gd – blue, Al – gray, Ge – green); red and black arrows denote Gd spins up and down in the ground-state magnetic configuration of GdAlGe. b) Calculated band structure of GdAlGe; bands with spins up and down are marked as red and blue, respectively.

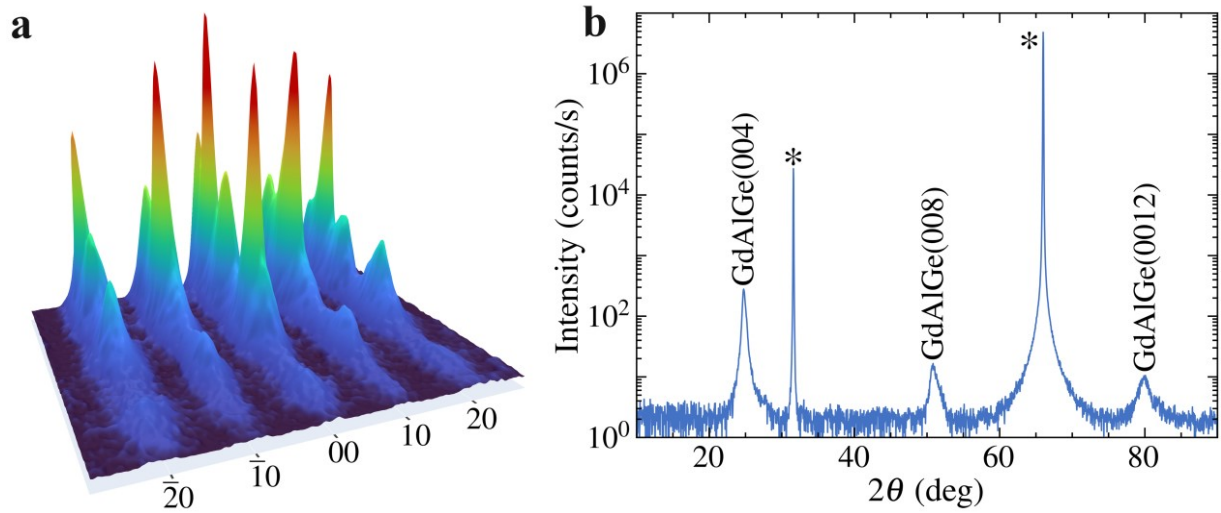

**Figure 2.** Structure of GdAlGe films. a) 3D RHEED image of 1 ML GdAlGe; the reflexes are marked by a pair of Miller indices for the basal plane. b) θ-2θ XRD scan of 90 ML GdAlGe; asterisks denote peaks from the Ge(001) substrate.

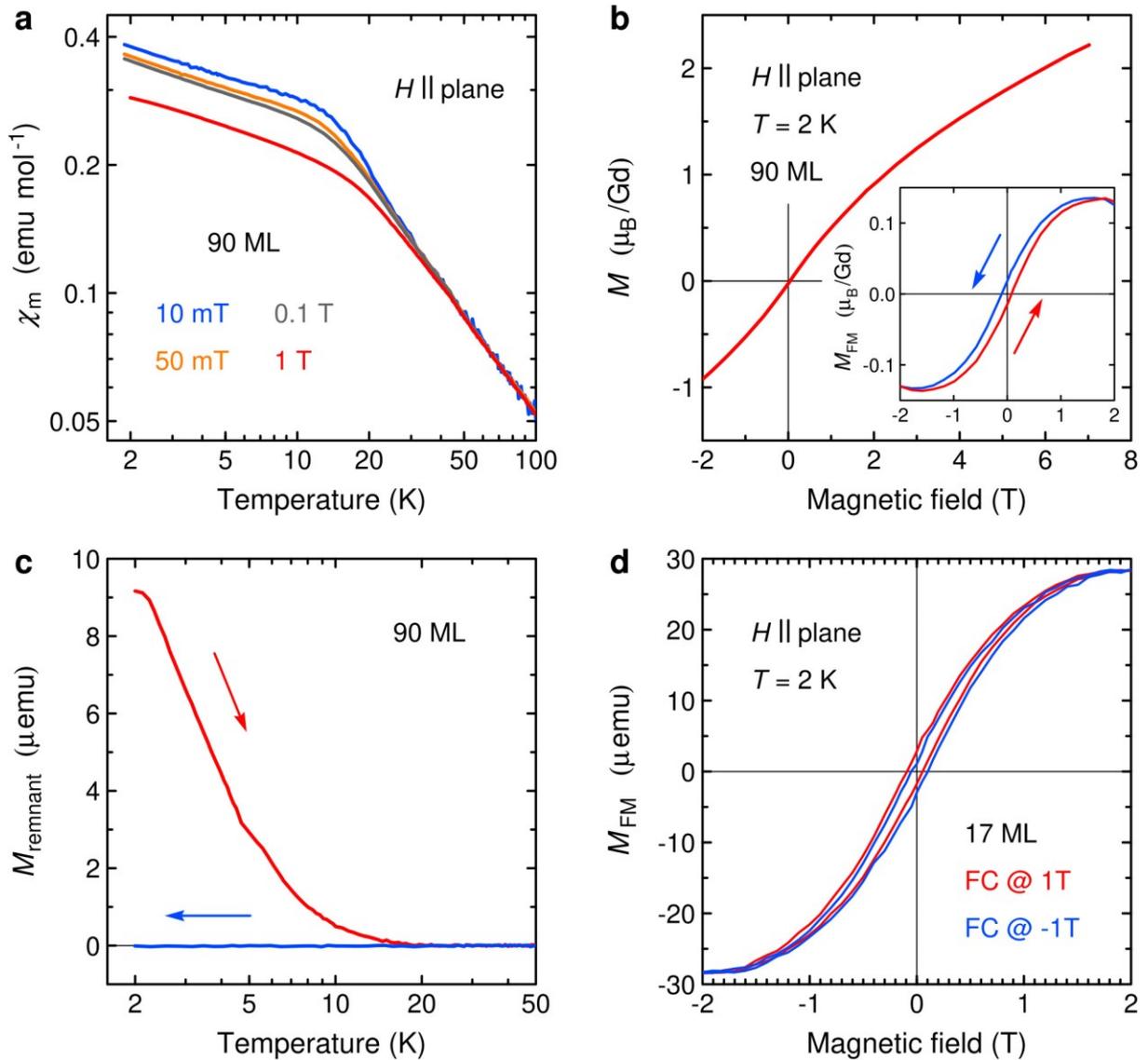

**Figure 3.** Magnetic properties of GdAlGe films. a) Temperature dependence of the molar magnetic susceptibility of 90 ML GdAlGe in in-plane magnetic fields 10 mT (blue), 50 mT (orange), 0.1 T (grey), and 1 T (red). b) Dependence of the magnetic moment per Gd atom on in-plane magnetic fields in 90 ML GdAlGe at 2 K; inset: Hysteresis in the dependence of the FM moment on in-plane magnetic fields in 90 ML GdAlGe at 2 K. c) Temperature dependence of the remnant moment after cooling 90 ML GdAlGe in an in-plane magnetic field 1 T. d) *M-H* hysteresis loops in 17 ML GdAlGe at 2 K following field cooling at 1 T (red) and -1 T (blue).

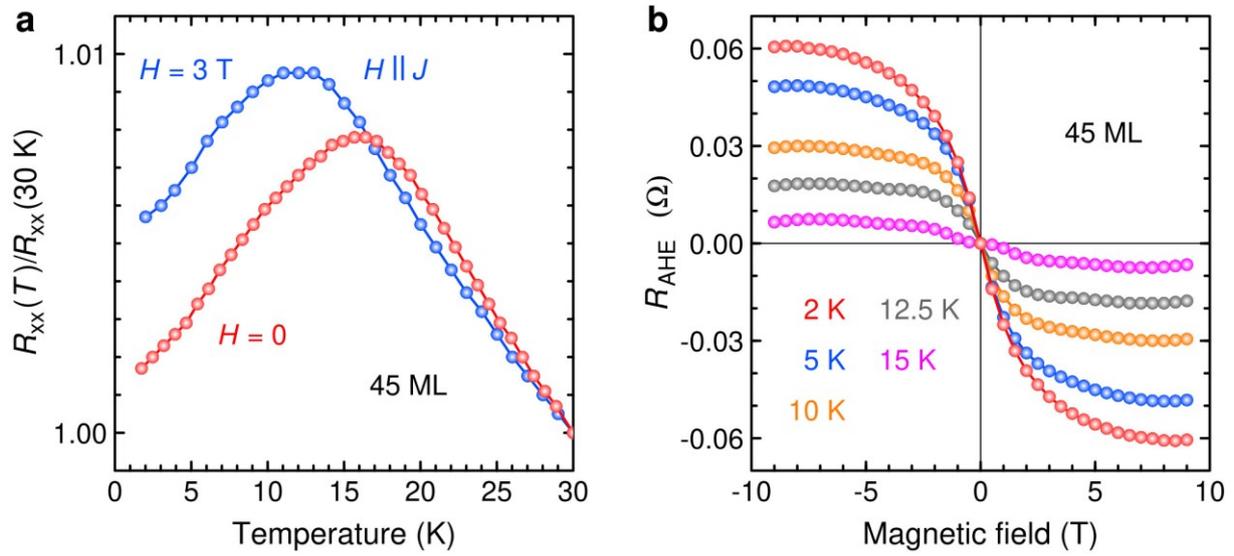

**Figure 4.** Lateral electron transport in 45 ML GdAlGe. a) Normalized temperature dependence of resistance in zero-magnetic field (red) and in a magnetic field 3 T parallel to the current. b) Non-linear (AHE) contribution to Hall resistance at 2 K (red), 5 K (blue), 10 K (orange), 12.5 K (gray), and 15 K (magenta).

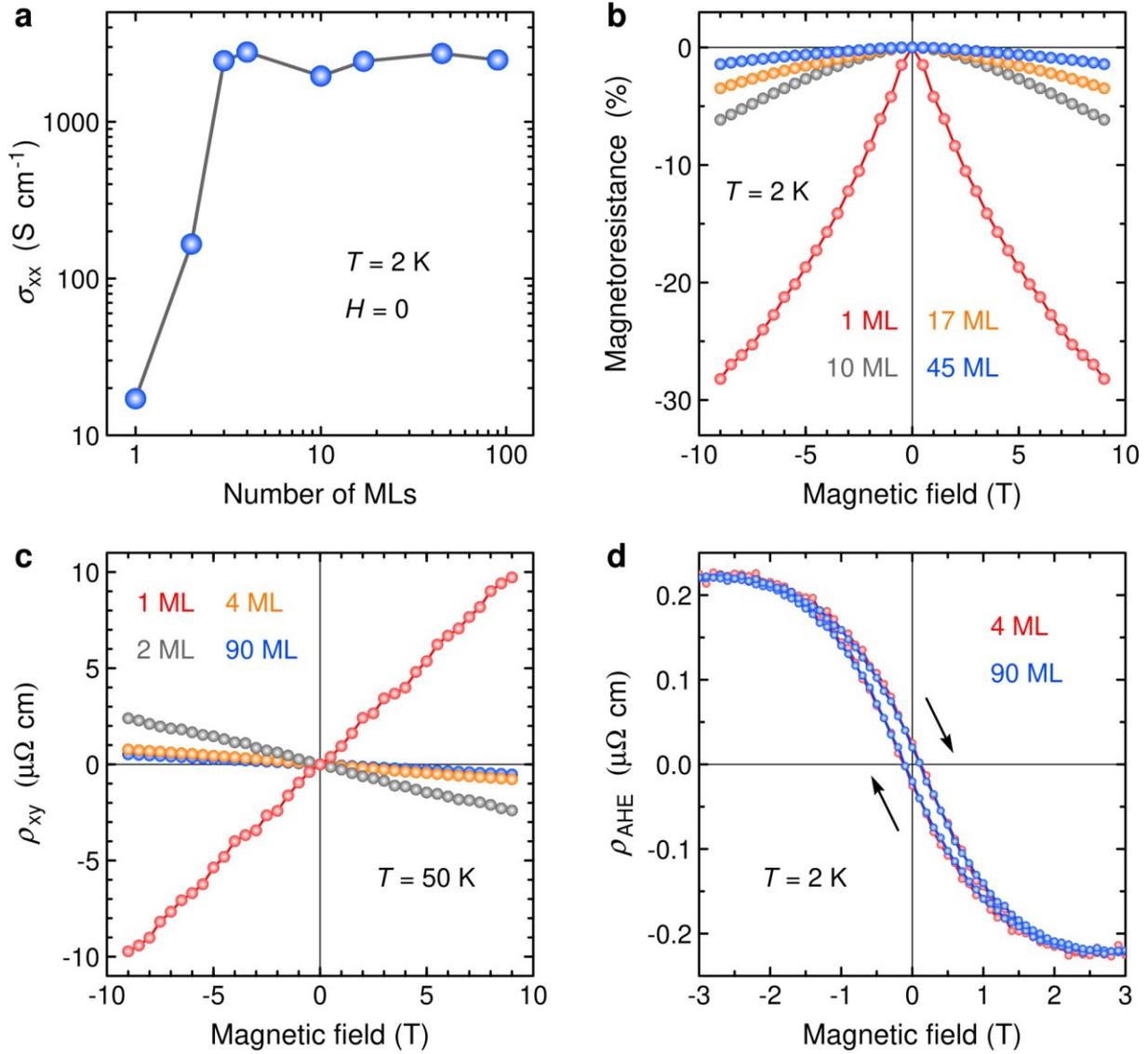

**Figure 5.** Thickness-dependent evolution of lateral electron transport in GdAlGe. a) Dependence of conductivity on the number of MLs at 2 K. b) MR for 1 ML (red), 10 ML (gray), 17 ML (orange), and 45 ML (blue) GdAlGe at 2 K in out-of-plane magnetic fields. c) Magnetic field dependence of Hall resistance for 1 ML (red), 2 ML (gray), 4 ML (orange), and 90 ML (blue) GdAlGe at 50 K. d) Magnetic field dependence of hysteresis in AHE resistance for 4 ML (red) and 90 ML (blue) GdAlGe at 2 K.

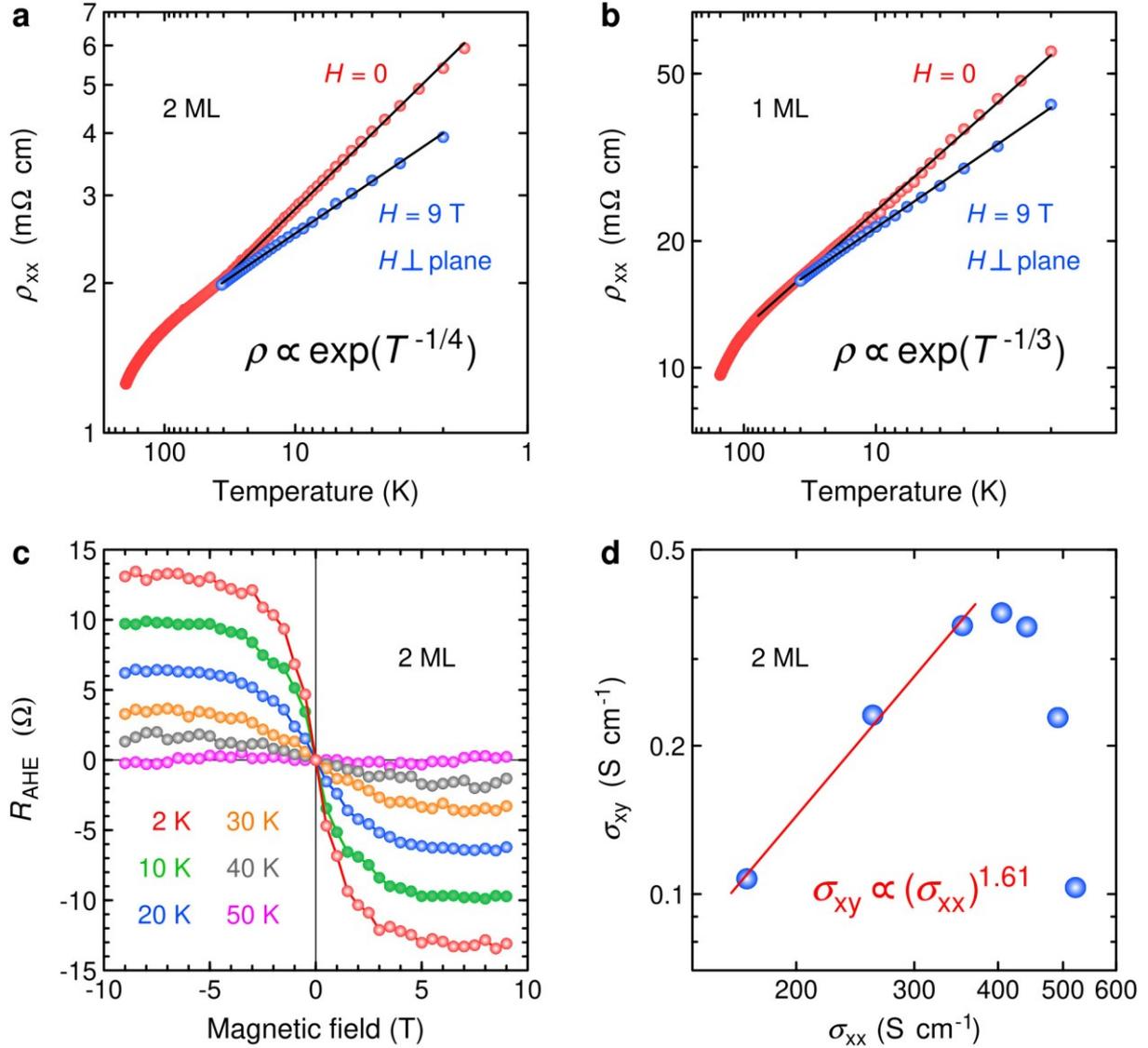

**Figure 6.** Lateral electron transport in ultra-thin films of GdAlGe. a) Temperature dependence of resistivity in 2 ML GdAlGe in zero magnetic field (red) and an out-of-plane magnetic field 9 T (blue); the fits (black lines) correspond to exponential dependence of resistivity on $T^{-1/4}$ (notice the use of a $T^{-1/4}$ x-scale and a logarithmic y-scale). b) Temperature dependence of resistivity in 1 ML GdAlGe in zero magnetic field (red) and an out-of-plane magnetic field 9 T (blue); the fits (black lines) correspond to exponential dependence of resistivity on $T^{-1/3}$ (notice the use of a $T^{-1/3}$ x-scale and a logarithmic y-scale). c) Non-linear (AHE) contribution to Hall resistance of 2 ML GdAlGe at 2 K (red), 10 K (green), 20 K (blue), 30 K (orange), 40 K (gray), and 50 K (magenta). d) Dependence of the AHE conductivity $\sigma_{xy}$ on the longitudinal conductivity $\sigma_{xx}$ in 2 ML GdAlGe. Red line shows the scaling relation $\sigma_{xy} \propto (\sigma_{xx})^{1.61}$.

# Supporting Information

**Competing Magnetic States in the Candidate Altermagnet GdAlGe**

*Oleg E. Parfenov, Dmitry V. Averyanov, Ivan S. Sokolov, Alexey N. Mihalyuk, Ivan A. Yakovlev, Oleg A. Kondratev, Alexander N. Taldenkov, Andrey M. Tokmachev, and Vyacheslav G. Storchak\**

**Content:**

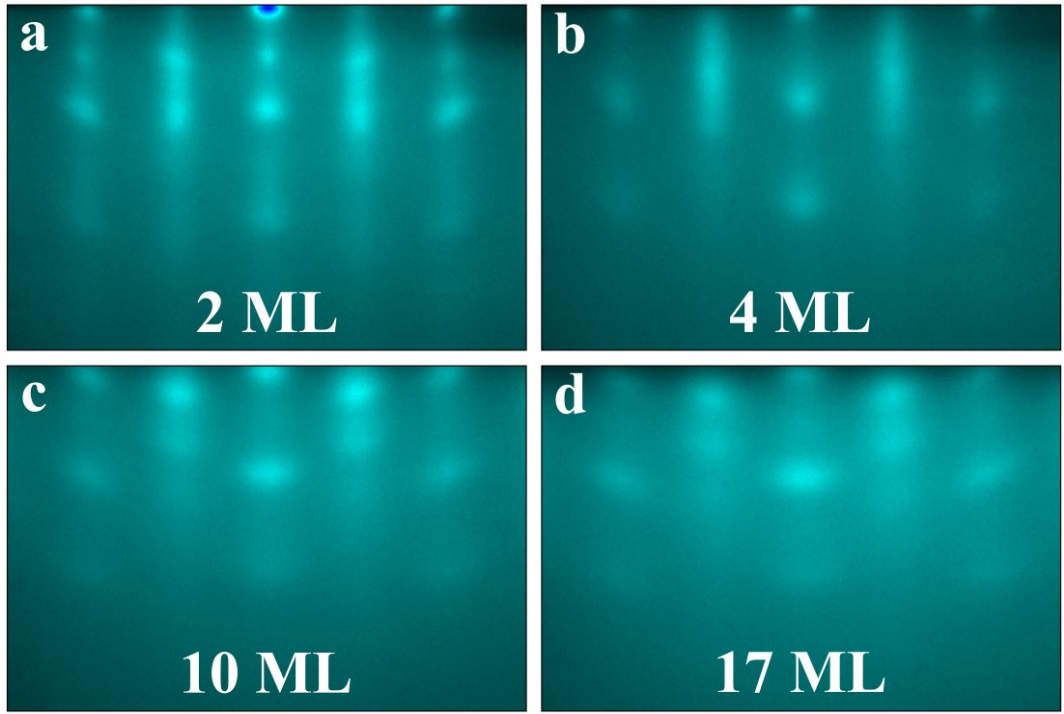

**Figure S1.** RHEED images of a) 2 ML, b) 4 ML, c) 10 ML, and d) 17 ML GdAlGe.

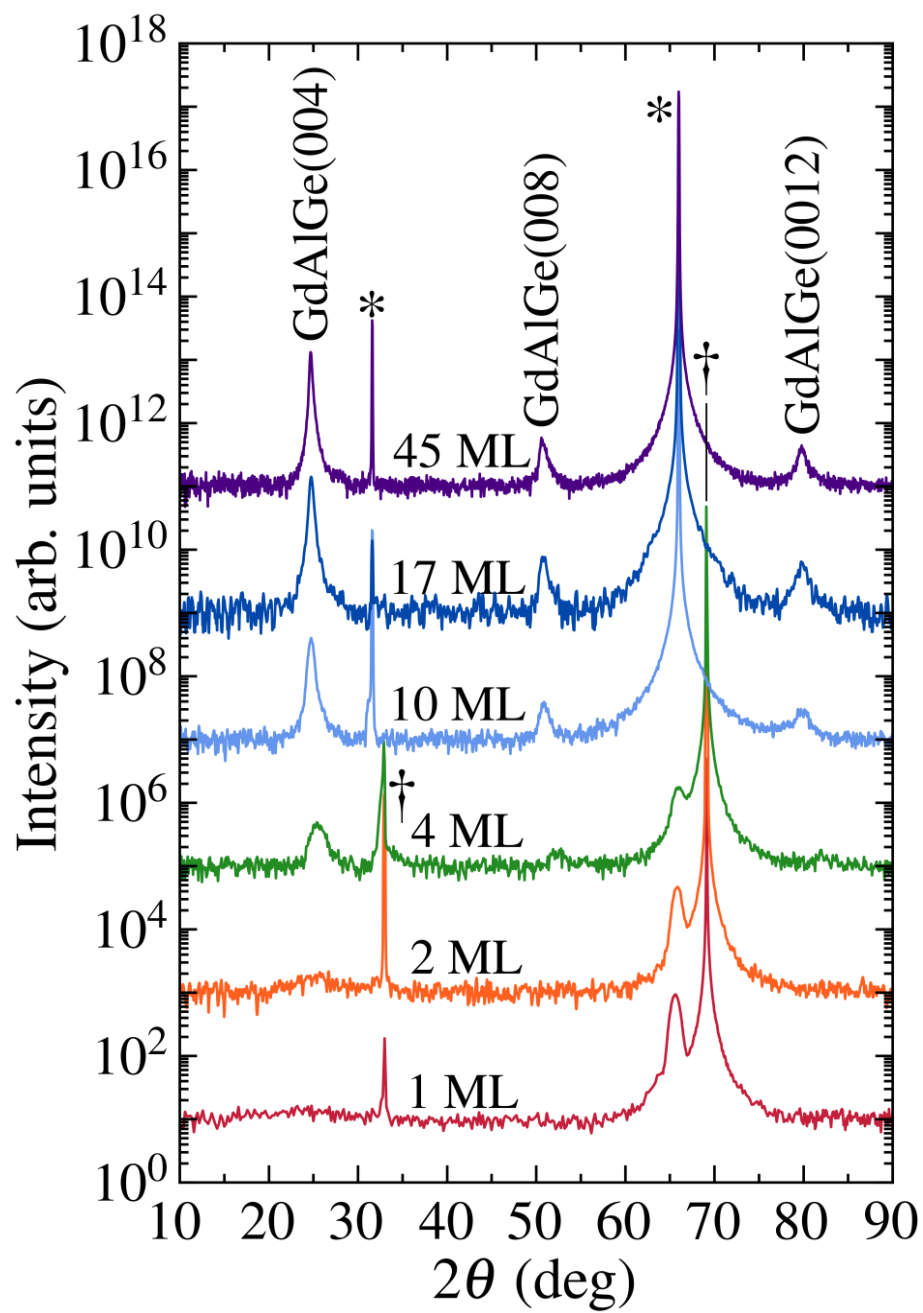

**Figure S2.** θ-2θ XRD scans of GdAlGe films: 1 ML (red), 2 ML (orange), 4 ML (green), 10 ML (light blue), 17 ML (dark blue), and 45 ML (purple). The curves are shifted to enhance the visibility of the peaks. Asterisks and daggers mark peaks from the Ge and Si substrates, respectively.

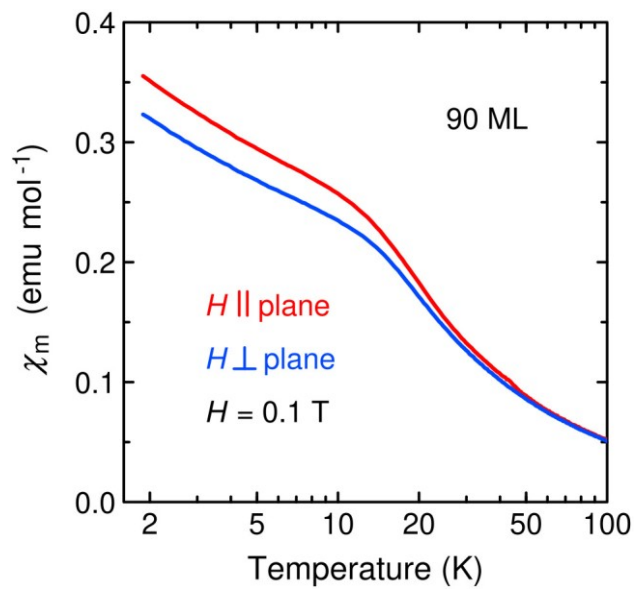

**Figure S3.** Temperature dependence of the molar magnetic susceptibility of 90 ML GdAlGe in in-plane (red) and out-of-plane (blue) magnetic fields of 0.1 T.

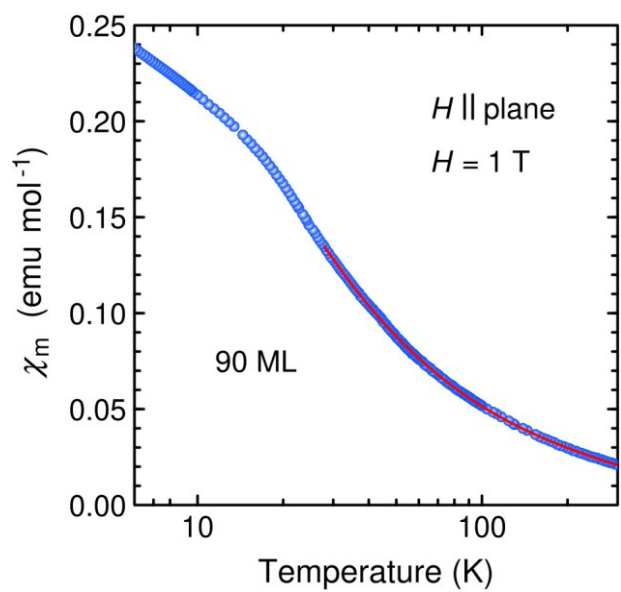

**Figure S4.** Temperature dependence of the molar magnetic susceptibility of 90 ML GdAlGe in an in-plane magnetic field 1 T (blue circles) and its fit by the Curie-Weiss law (red line).

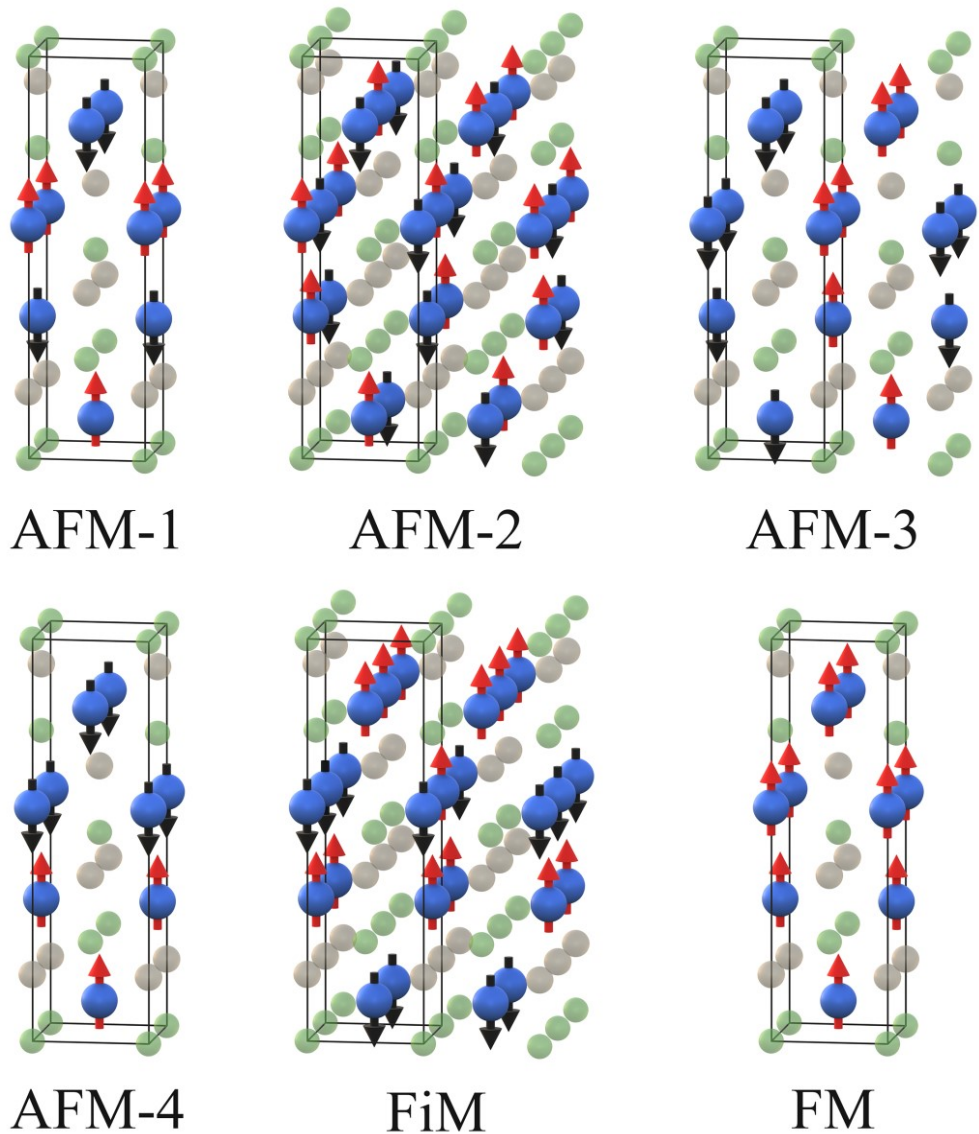

**Figure S5.** Magnetic configurations of GdAlGe employed in DFT calculations – magnetic unit cells for 4 antiferromagnetic configurations (AFM-1, AFM-2, AFM-3, and AFM-4); one ferrimagnetic configuration (FiM), and one ferromagnetic configuration (FM). Red and black arrows denote Gd spins up and down, respectively.

| Magnetic configuration | AFM-1 | AFM-2 | AFM-3 | AFM-4 | FiM | FM |
|---|---|---|---|---|---|---|
| Relative energy (meV) | 0.0 | 2.0 | 5.0 | 17.4 | 0.9 | 45.7 |

**Table S1.** DFT calculated energies of GdAlGe for the 6 magnetic configurations presented in Figure S5. The energies (per formula unit) are relative to the energy of the altermagnetic AFM-1 configuration.

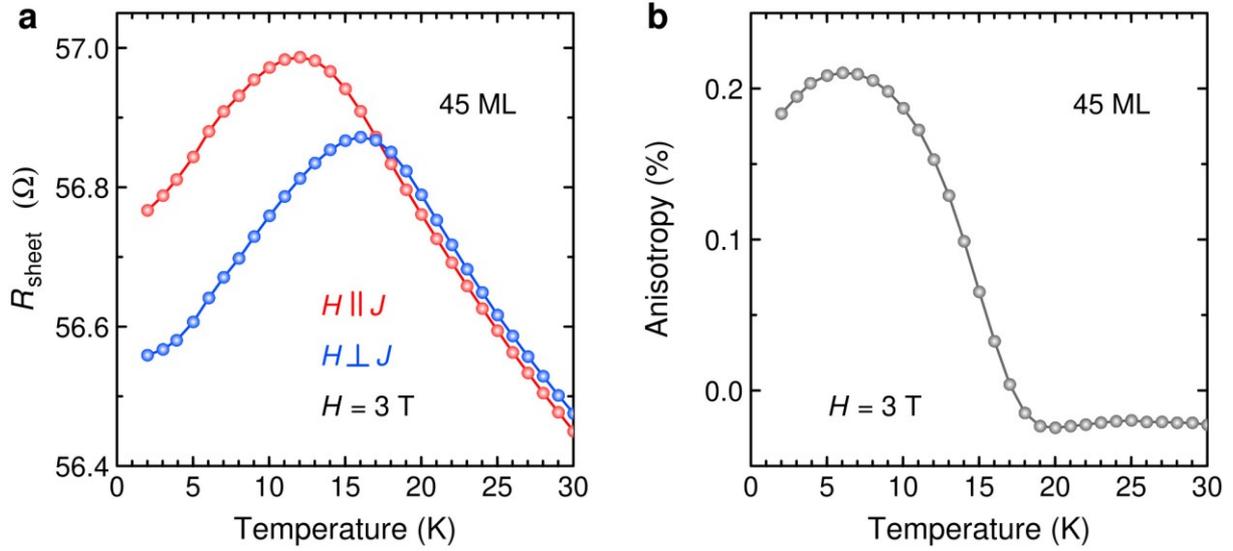

**Figure S6.** a) Temperature dependence of sheet resistance of 45 ML GdAlGe in an in-plane magnetic field 3 T parallel (red) and perpendicular (blue) to the current. b) Temperature dependence of anisotropy in the resistance of 45 ML GdAlGe in an in-plane magnetic field 3 T, calculated as the difference between resistances in magnetic fields parallel and perpendicular to the current normalized to their average value.

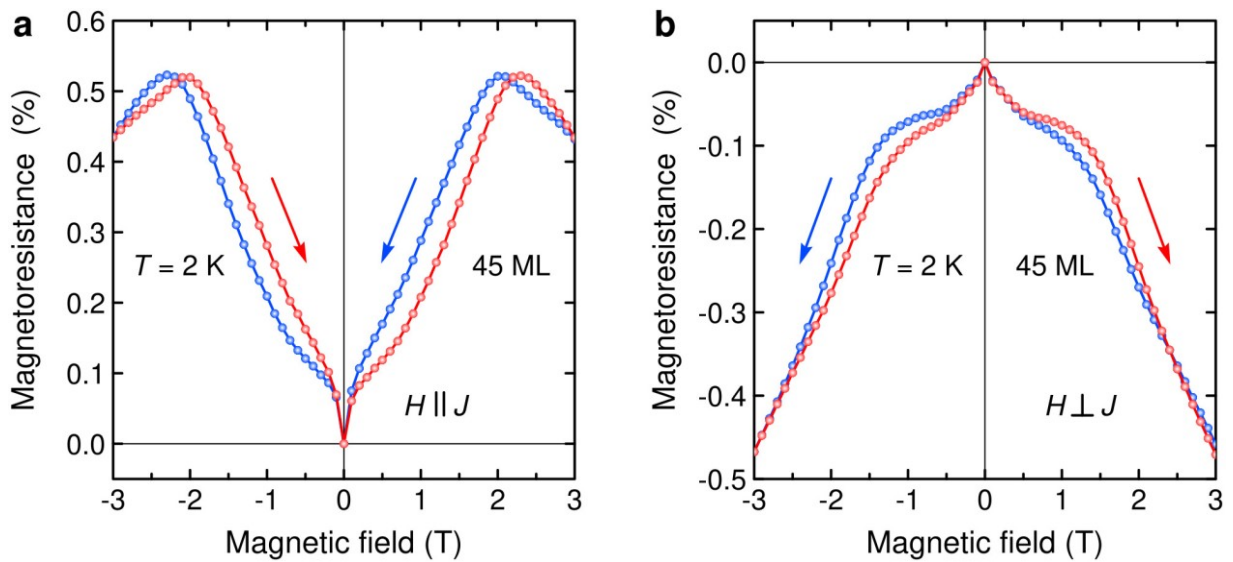

**Figure S7.** Hysteresis in MR of 45 ML GdAlGe at 2 K for in-plane magnetic fields a) parallel and b) perpendicular to the current.

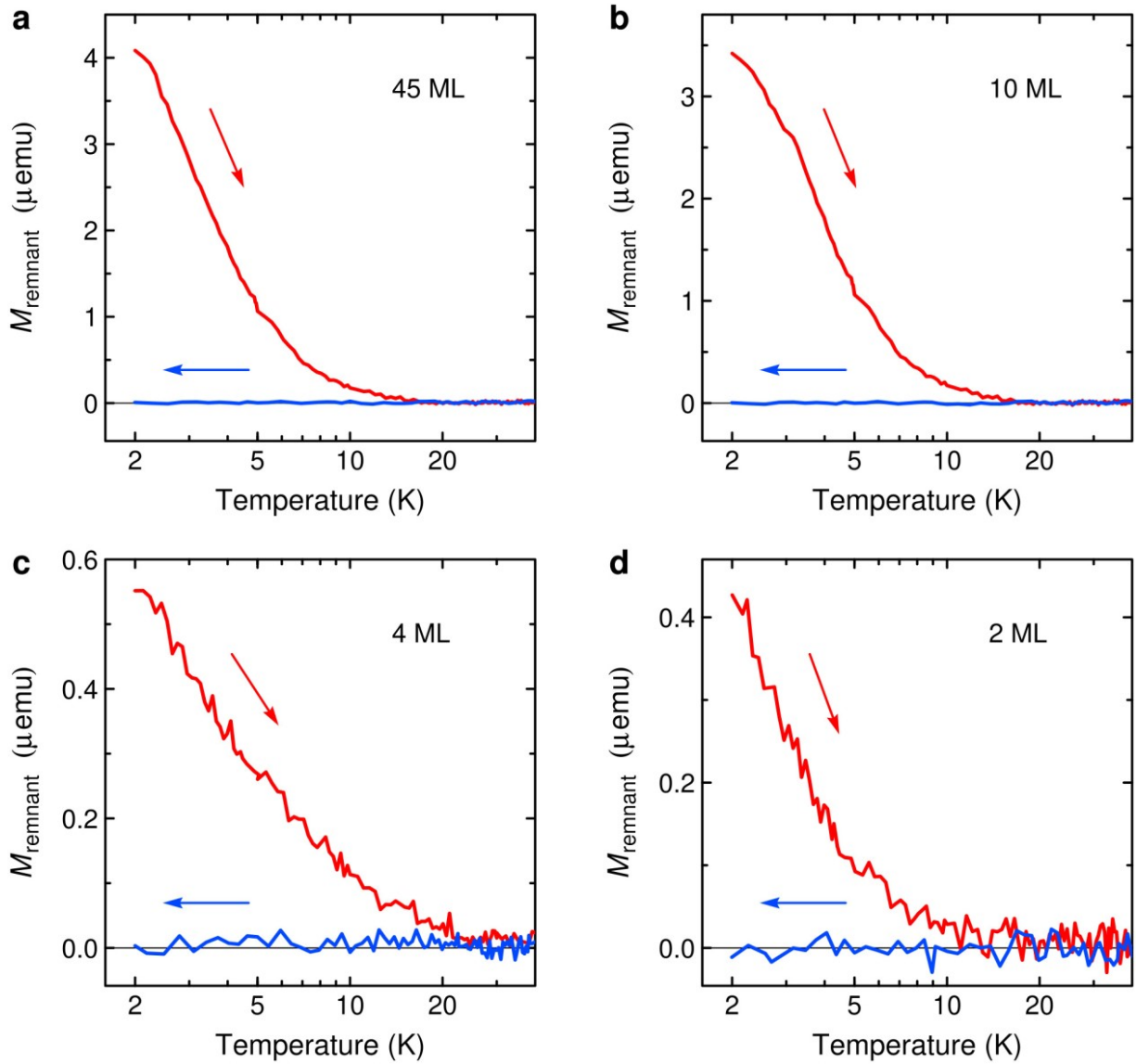

**Figure S8.** Temperature dependence of the remnant moment in a) 45 ML, b) 10 ML, c) 4 ML, and d) 2 ML GdAlGe after cooling in an in-plane magnetic field 1 T.

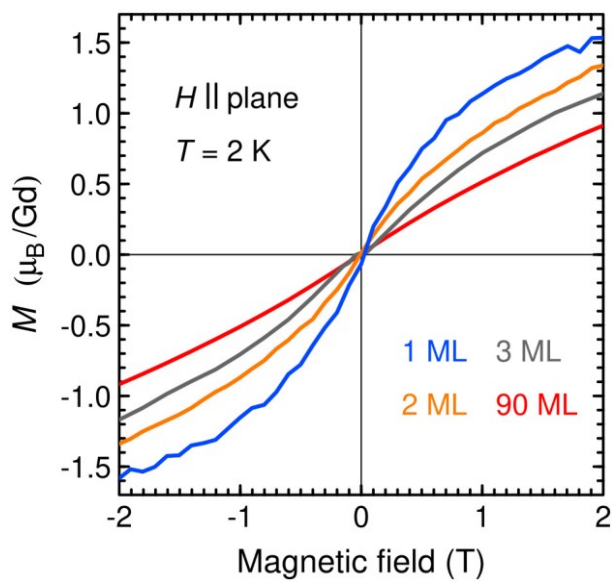

**Figure S9.** In-plane magnetic field dependence of the magnetic moment per Gd atom in 1 ML (blue), 2 ML (orange), 3 ML (gray), and 90 ML (red) GdAlGe at 2 K.

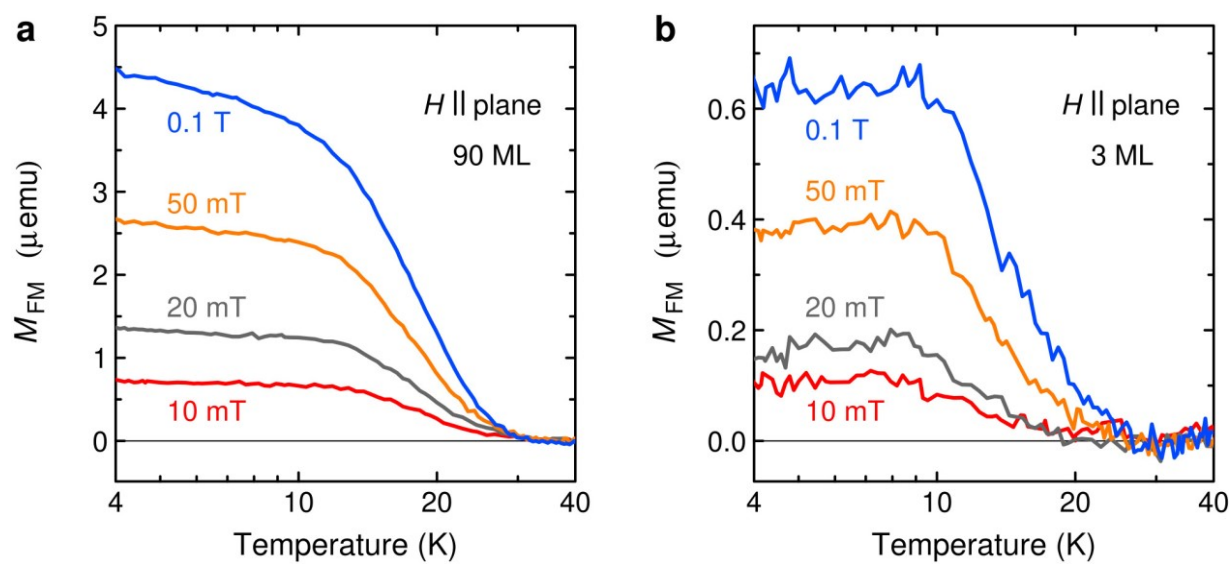

**Figure S10.** Temperature dependence of the FM moment in a) 90 ML and b) 3 ML GdAlGe in in-plane magnetic fields of 10 mT (red), 20 mT (gray), 50 mT (orange), and 0.1 T (blue).

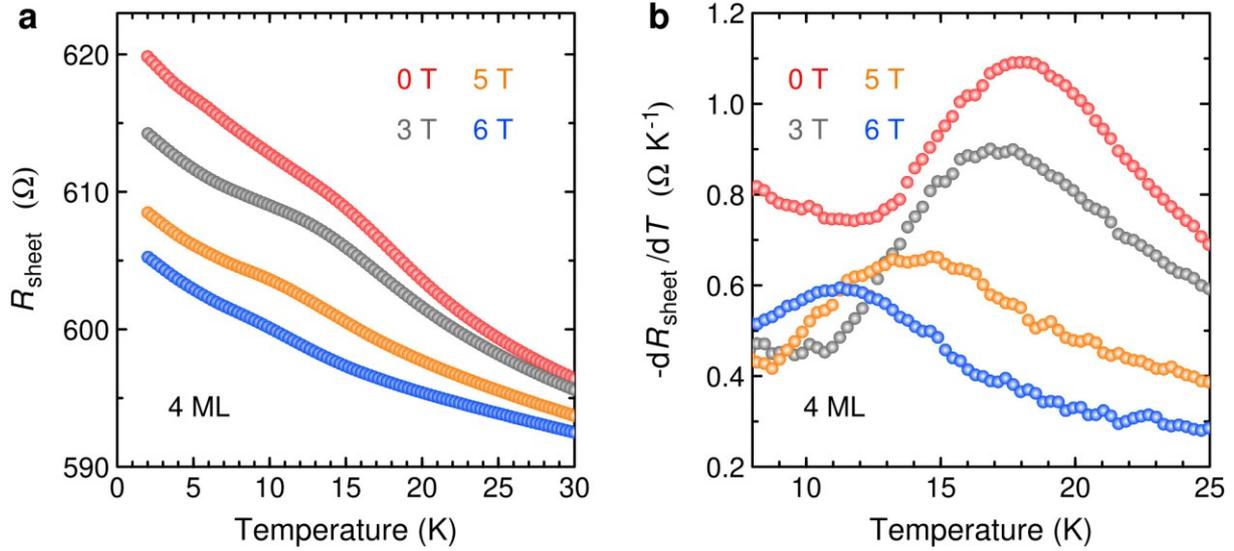

**Figure S11.** a) Temperature dependence of sheet resistance and b) its derivative in 4 ML GdAlGe in zero magnetic field (red) and out-of-plane magnetic fields 3 T (gray), 5 T (orange), and 6 T (blue).

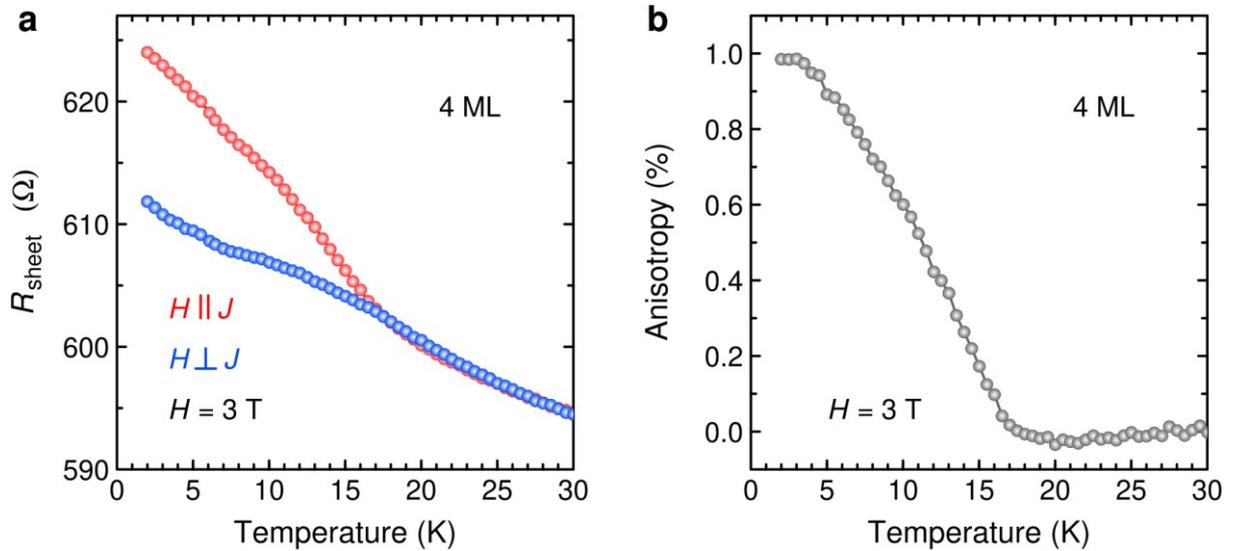

**Figure S12.** a) Temperature dependence of sheet resistance of 4 ML GdAlGe in an in-plane magnetic field 3 T parallel (red) and perpendicular (blue) to the current. b) Temperature dependence of anisotropy in the resistance of 4 ML GdAlGe in an in-plane magnetic field 3 T, calculated as the difference between resistances in magnetic fields parallel and perpendicular to the current normalized to their average value.

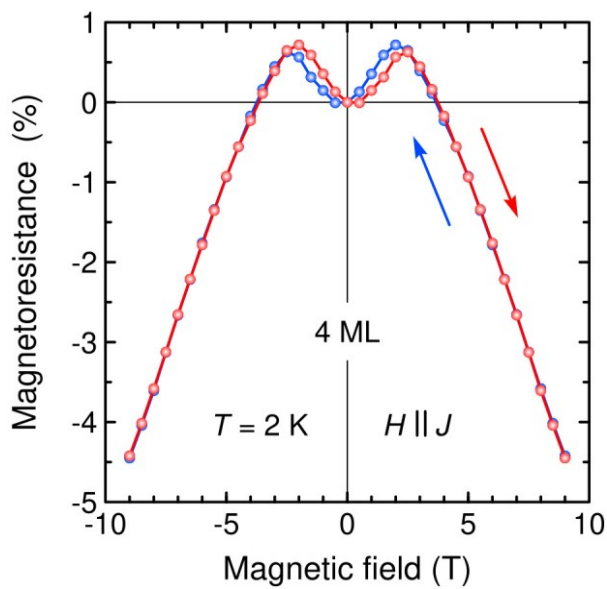

**Figure S13.** Hysteresis in MR of 4 ML GdAlGe at 2 K in magnetic fields parallel to the current.

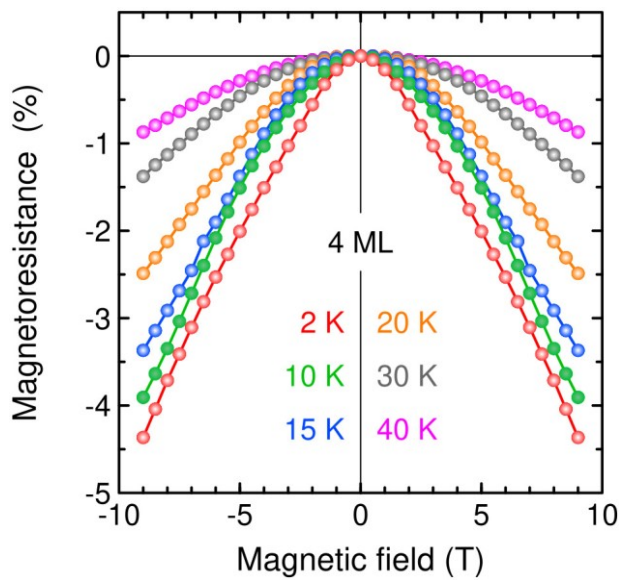

**Figure S14.** Out-of-plane magnetic field dependence of MR in 4 ML GdAlGe at 2 K (red), 10 K (green), 15 K (blue), 20 K (orange), 30 K (gray), and 40 K (magenta).

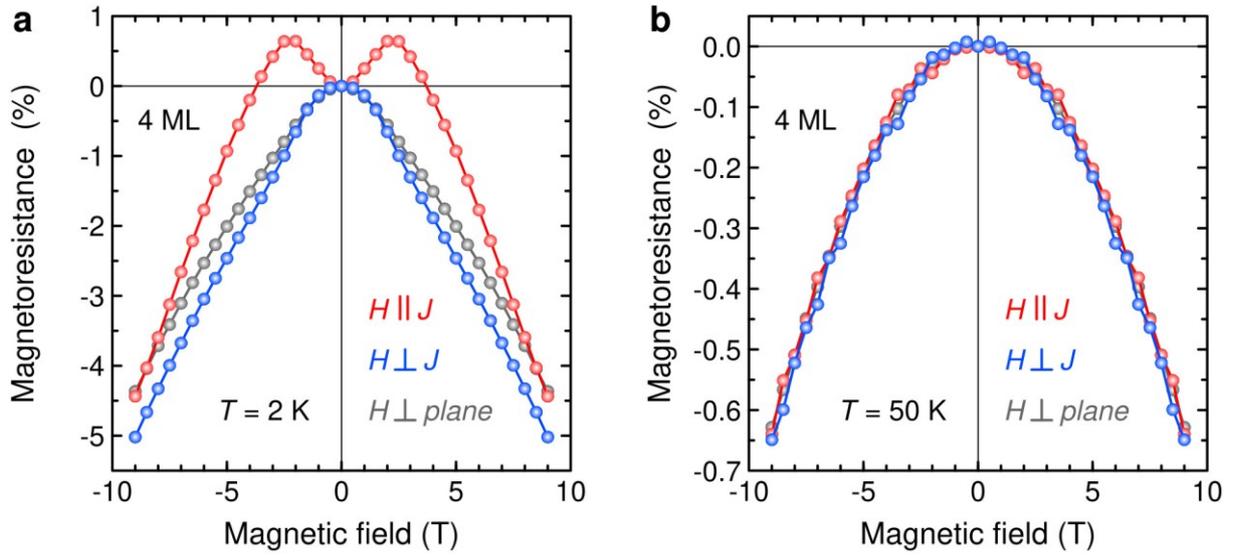

**Figure S15.** MR in 4 ML GdAlGe at a) 2 K and b) 50 K for out-of-plane magnetic fields (gray) and in-plane magnetic fields parallel (red) and perpendicular (blue) to the current.

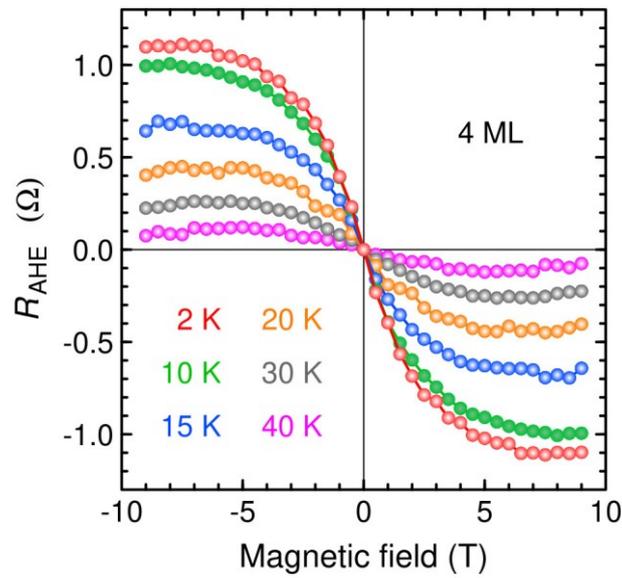

**Figure S16.** Non-linear (AHE) contribution to Hall resistance of 4 ML GdAlGe at 2 K (red), 10 K (green), 15 K (blue), 20 K (orange), 30 K (gray), and 40 K (magenta).

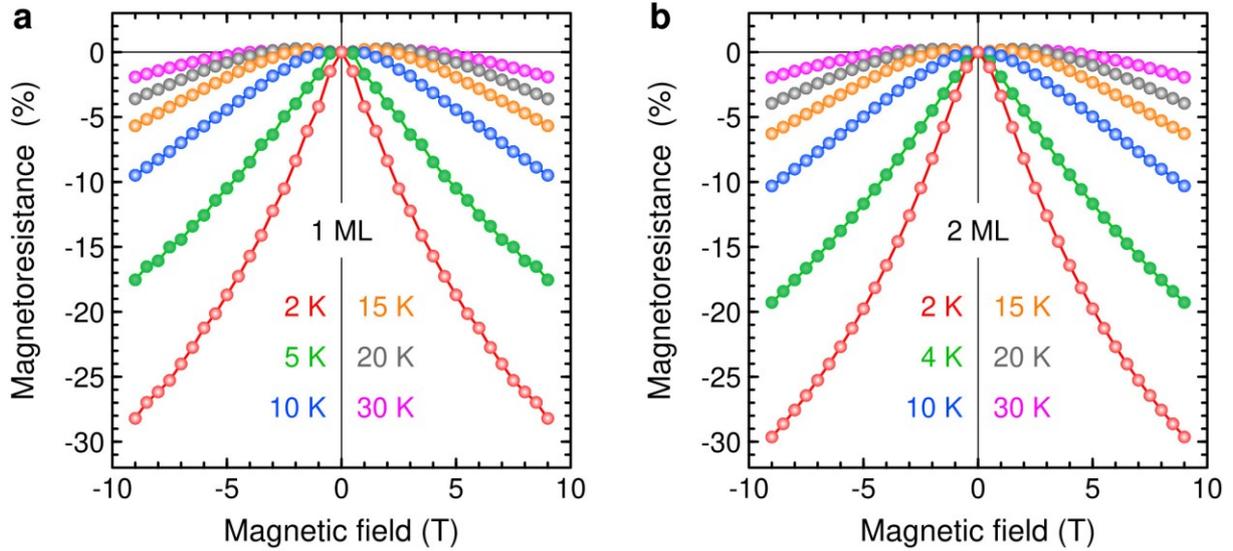

**Figure S17.** a) Out-of-plane magnetic field dependence of MR in 1 ML GdAlGe at 2 K (red), 5 K (green), 10 K (blue), 15 K (orange), 20 K (gray), and 30 K (magenta). b) Out-of-plane magnetic field dependence of MR in 2 ML GdAlGe at 2 K (red), 4 K (green), 10 K (blue), 15 K (orange), 20 K (gray), and 30 K (magenta).

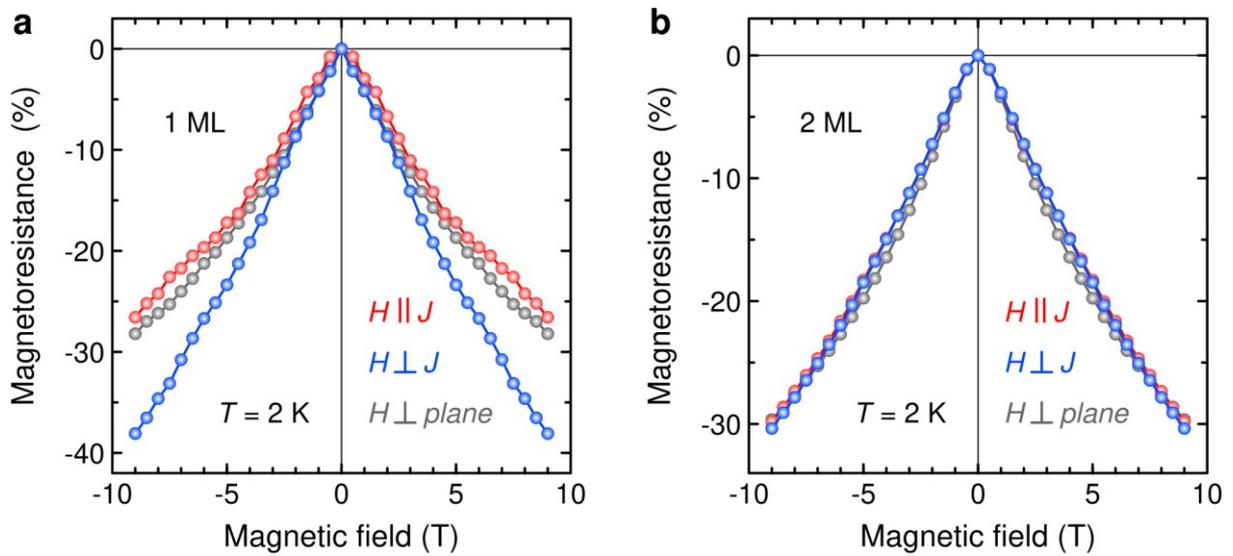

**Figure S18.** MR in a) 1 ML and b) 2 ML GdAlGe for out-of-plane magnetic fields (gray) and in-plane magnetic fields parallel (red) and perpendicular (blue) to the current at 2 K.